\documentclass[journals]{IEEEtran}
\IEEEoverridecommandlockouts

\makeatletter
\def\markboth#1#2{\def\leftmark{\@IEEEcompsoconly{\sffamily}\MakeUppercase{\protect#1}}%
\def\rightmark{\@IEEEcompsoconly{\sffamily}\MakeUppercase{\protect#2}}}
\makeatother

\usepackage[utf8x]{inputenc}
\usepackage[english]{babel} 
\usepackage{ucs} 
\usepackage{amsmath} 
\usepackage{amsfonts} 
\usepackage{amssymb} 
\usepackage{amsthm}
\usepackage{array}
\usepackage{verbatim}
\usepackage{listings}
\usepackage{stfloats}

\usepackage{algorithm} 
\usepackage{algorithmic} 
\usepackage{url} 
\usepackage{enumerate}
\usepackage{multirow}
\usepackage{longtable}

\usepackage{caption}
\usepackage{subcaption}  
\usepackage[dvipdf,pdftex]{graphicx}
\usepackage{epstopdf}
\usepackage{graphicx}
\usepackage{bm}
\usepackage{units}
\usepackage{color}
\usepackage{hhline}
\def\beq{\begin{equation}}
\def\eeq{\end{equation}}
\def\beqa{\begin{eqnarray}}
\def\eeqa{\end{eqnarray}}
\def\beqan{\begin{eqnarray*}}
\def\eeqan{\end{eqnarray*}}

\title{\Large Provisioning Low Latency, Resilient Mobile Edge Clouds for 5G}

\author{
    \IEEEauthorblockN{
	    Russell Ford,~\IEEEmembership{Student Member,~IEEE},
	    Ashwin Sridharan,~\IEEEmembership{Member,~IEEE},
	    Robert Margolies,~\IEEEmembership{Member,~IEEE},\\
	    Rittwik Jana,~\IEEEmembership{Member,~IEEE},
	    Sundeep Rangan,~\IEEEmembership{Fellow,~IEEE}
	    }
} 

\begin{document} 

\maketitle
%
%
%

\begin{abstract}
Network virtualization and SDN-based routing allow carriers to flexibly configure
their networks in response to demand and unexpected network disruptions. However,
cellular networks, by nature, pose some unique challenges because of 
user mobility and control/data plane partitioning, which calls for new architectures
and provisioning paradigms. In this paper, we address the latter part by devising
algorithms that can provision the data plane to create a distributed Mobile Edge Cloud (MEC), which provides 
opportunities for lower latencies and increased resilience (through placement of network functions 
at more distributed datacenter locations) \textit{and} accounts for service disruption that could be incurred
because of user mobility between the service areas of different datacenters. Through evaluations
with topology and traffic data from a major carriers's network, we show that,
compared to static, centralized networks, careful virtualized provisioning can yield
significant savings in network costs while still minimizing service disruption due to mobility.
We demonstrate that up to a 75\% reduction in redundant datacenter capacity over the operator's current topology 
(while achieving the same level of resilience)
is possible by distributing load over many mobile cloud datacenters.
\end{abstract}


\section {Introduction}

Current cellular core networks (CNs) are highly centralized and composed of monolithic Network Elements (NEs) placed at a small number of core datacenters (DCs). In typical deployments of the 4G Evolved Packet Core (EPC), these NEs have, until recently, taken the form of high-capacity, high-reliability ``Big Iron'' appliances, which are expensive, complex to manage and are often statically-provisioned to handle the peak demand. Consequently, a significant amount of redundant EPC infrastructure is required to provide resilience and fault-tolerance, since, if a node fails, a secondary node must be on standby in order to restore the sessions for large number of users. Clearly, this form of ``1+1'' redundancy is a costly prospect for operators, especially as they now confront the anticipated 1000-fold surge in data traffic and 100-fold increase in number of connected devices~\cite{popovsk2013eu}.

The sparse geographic placement of core NEs and Internet Exchange Points (IXPs), where operators peer with other networks, also poses a dilemma for offering low-latency services over the top of mobile networks, since traffic must flow through the centralized IXP routers and NEs before proceeding to the edge. Achieving the near-instantaneous user experience required by many emerging interactive, real-time \textit{Tactile Internet} applications may require end-to-end (E2E) latencies below 10 milliseconds. Additionally, new mission-critical 5G use cases, such as monitoring and control of factory automation and self-driving cars, presents a need for latencies of 1 ms or less \cite{popovsk2013eu,5gtactile}. However, due to the potential for long routes between Internet servers and the access network, the target latency for many of these applications cannot reliably be met in the current 4G core \cite{SMORE:14}.

Hence, today's cellular core networks can be characterized as highly centralized, inelastic, inflexible and redundant, as well as inherently ill-suited for providing low-latency 5G services. However, Software Defined Networking (SDN) and Network Function Virtualization (NFV) are two key technologies, which have the potential to streamline the next-generation network architecture. 
Virtualization of the CN stack through NFV enables lightweight, scalable software instances of Virtual Network Functions (VNFs) to be deployed on commodity servers. Since the processing resources required by a VNF can be scaled to match its specific load, many of the costs of managing and provisioning resources for network functions can be mitigated. This makes it possible to deploy core VNFs and even application-layer services on general-purpose hardware at many smaller DCs located closer to the edge, toward the realization of a Mobile Edge Cloud (MEC). MEC networks, enabled by SDN/NFV, would offer a number of opportunities for operators to optimize their networks and enhance service for end-users. 

\begin{itemize}
	\item \textbf{Latency:} Distributing network functions and applications to edge sites would allow operators to offer new value-added services with latency requirements that cannot currently be met via over-the-top delivery \cite{popovsk2013eu,5gtactile,SMORE:14}.
	\item \textbf{Resiliency:} Since load can be distributed amongst hundreds or thousands of virtual NEs at many smaller DC sites, the effect of node failures can be mitigated and the amount of redundant capacity required to provide the same degree of resilience is reduced.
	\item \textbf{Reduced backbone traffic:} Caching content and hosting applications closer to the edge can be expected to reduce the cost of transporting traffic over long distances from centralized IXPs to the access network. The need for high-capacity backbone routers may also be alleviated.
	\item \textbf{Elastic provisioning and intelligent routing} provided by NFV and SDN, respectively, allow traffic to be dynamically redistributed across servers and DC sites in response to fluctuations in demand or failures. Resources can thus be allocated in the most cost-effective way based on current network conditions.
\end{itemize}

While SDN/NFV offers many benefits, architecting the MEC presents numerous challenges due
to the need to support mobility as well as other functions that are uniquely required by mobile networks. As network functions and services are pushed out closer to the access edge, user movement will trigger more frequent session migrations between serving DC nodes. Each migration results in increased signaling load on network controllers and can cause service interruptions, which adversely affect user experience. 
In addition to experiencing more migrations, the MEC also has reliability concerns, as it is reasonable to expect software running on general-purpose servers will be inherently less reliable compared to highly-optimized network appliances. Although individual commodity servers may be less reliable, we argue that distribution of load across many edge nodes makes it possible for networks, as a whole, to be more reliable and at less cost than current EPCs. 

Operators must therefore provision resources, assign user sessions to serving DCs and route traffic while taking into account diverse requirements for latency, fault tolerance, cost of DC capacity, link bandwidth, and user state migrations. In this work, we investigate this problem in the context of SDN/NFV-based MEC networks. While there is a body of recent related work on mobile cloud architecture and optimization \cite{Kempf:12,Said:13,SoftCell:13,Basta:13,MobileFlow:13,SDMA_SDN:13,Basta:14,Design_5G_CommMag:14,SoftMoW:14,SMORE:14,Baumgartner:15,KLEIN:16}, this work is the first to incorporate each of these real-world costs and design constraints and analyze the key tradeoffs between latency, resilience and cost of service migration.

The remainder of the paper is organized as follows. We begin in Section~\ref{sec:mec_model} by introducing the MEC resource allocation problem and system model. In Section~\ref{sec:mec_opt}, we formulate the optimization problem and present a set of heuristics, which can be used to jointly associate users with serving DCs, allocate resources for DCs and links, and route traffic on a large scale in order to satisfy the set of operator and end-user requirements. 
Finally, we apply the proposed techniques to a major operator's network topology, which includes DC and base station (BS) sites from a large metropolitan region in the United States and data on current traffic demands and inter-cell handover rates. We demonstrate that, in addition to being able to flexibly meet the latency requirements of 5G applications, a 75\% reduction in redundant DC capacity as well as significant reductions in backbone link bandwidth are possible by distributing demand across smaller mobile cloud DCs. 

\section{Problem Definition and System Model}
\label{sec:mec_model}
\subsection{Problem Definition}
In this work, we are concerned with the problem of optimally provisioning resources in the MEC and associating users with serving DCs to service the maximal end-user demand. Most data traffic delivered over current cellular core networks originates from the Internet and is routed between the IXP router, located at one of a small number of fixed DC sites, and the BS where the user is connected. However, envisioned MEC architectures make it possible for application VMs, such as gaming servers, voice/video telephony servers and CDN nodes, to be instantiated at many DC sites where cloud computation and storage capacity is available. Application VMs can hence be hosted internally to the operator's network. In this work, we choose to focus on optimizing MEC-based core networks specifically for these types of operator-hosted applications.

We wish to route flows and provision resources on a periodic basis to minimize network costs while satisfying all E2E latency requirements. Such a capability is useful for operators, who must predict hourly peak demand and mobility patterns based on past trends to guarantee sufficient resources are brought online to prevent the network from becoming overloaded. Similarly, resources may be de-provisioned and scaled down when the network is lightly loaded to save on operational costs (i.e., OpEx). In MEC networks, a mobile cloud controller will take traffic and handover (HO) volume predictions and determine which servers, VMs and line cards should be brought up or down ahead of time such that the minimum cost is incurred and the greatest possible demand is able to be serviced. 

While we recognize that next-generation core networks will rely on such technologies as SDN and NFV, the specifics of the architecture and protocol are, as yet, unknown. Therefore, we attempt to make the system model formulation and algorithms introduced in the following sections as agnostic to the design of future protocols as possible.


\subsection{System Model}
We consider a network model that is represented by the topology graph $\mathcal{G} = (\mathcal{V},\mathcal{E})$, where $\mathcal{V} = \{\mathcal{V}^{\rm dc},\mathcal{V}^{\rm bs}\}$ contains the set of all DC nodes $\mathcal{V}^{\rm dc}$ and BS nodes $\mathcal{V}^{\rm bs}$ in the operator's network and $\mathcal{E}$ is the set of fiber links interconnecting them. 
Among these sites are the operator-controlled core DCs and central office (CO) sites, the latter of which we shall refer to as \textit{leaf DCs} since they serve as the point of termination for the backhaul links connecting eNodeB base stations to the core network. For later convenience, we define an \textit{access/backhaul network sub-graph} $\mathcal{G^{\rm bs}} = (\mathcal{V^{\rm bs}},\mathcal{E^{\rm bs}})$, which is composed of the BSs, leaf DCs and backhaul links. 



Let $d$ denote an aggregate traffic flow for a number of users rooted at the serving BS $i_d \in \mathcal{V^{\rm bs}}$. Flow $d$ is served by DC $v_d \in \mathcal{V^{\rm dc}}$, which can be any DC within the latency budget $W_d$ of the flow. Let $h_d(t)$ be the total traffic demand associated with $d$ at time $t$. This demand can be divided and routed along multiple paths $p_d \in P_{i_d,v_d}$ in either the downlink or uplink direction, where $P_{i_d,v_d}$ is the set of paths between $i_d$ and $v_d$ having latencies $w_{p_d} \le W_d$. The latency requirement thus restricts the set of candidate serving DCs to those within a certain path latency from a given BS. We refer to this set of candidate DCs as the \textit{service area} $DC(i)$ for the BS $i$. Similarly, we define the service area $SA(v)$ of a DC $v$ as the set of BS nodes that are reachable from $v$ within a stated latency budget. It is possible to define different SAs depending on the latency constraints of various service types.



Additionally, links $e \in \mathcal{E}$ and DCs $v \in \mathcal{V^{\rm dc}}$ have bandwidths and processing capacity $B^{\rm link}_e$ and $C^{\rm dc}_v$, respectively. Let $b^{\rm dc}$ be the cost per bit of DC processing capacity and $b^{\rm link}_e$ be the cost of bandwidth on link $e$, which we assume to be proportional to the length of the fiber link. We therefore write $b^{\rm link}_e = b^{\rm link} D_e$, where $b^{\rm link}$ is the cost per bit per mile and $D_e$ is the distance of $e$ in miles. 

\paragraph*{Modeling Handover and Service Area Relocation}
Each BS $i$ will see some volume of HOs $\lambda_{i,j}(t)$ to other BSs $j \in \mathcal{N}(i)$, where $\mathcal{N}(i)$ is the set of neighboring BSs to $i$. Like data traffic, the number of HOs will vary over the course of the day depending on the number of active UEs and their mobility patterns. We note that, since the access network topology is fixed and UE mobility is not controlled by the operator, the HO volume experienced by the network is considered a given. 
The operator does, on the other hand, exercise control over which serving DC a user flow associates with (upon attaching to the network, for instance), which will determine when migrations will occur. When a user served by DC $v$ undergoes a HO from source BS $i$ to target BS $j$ that is outside $SA(v)$ (that is, the UE's current serving DC no longer meets its latency constraint), its session must be migrated to a new DC that is inside the service area $DC(j)$ of the target BS and is able to satisfy its latency budget. This procedure is known as Service Area Relocation (SAR). As user Quality of Experience (QoE) may be affected by frequent SAR due to increases in service interruption time and a higher chance of migration failure, it is reasonable to assume that operators will want to limit the rate of SAR experienced by users.

\paragraph*{Modeling Resilience}
Resilience and fault tolerance have been investigated extensively in previous work. We choose to adopt a simple model for resilience where only failures of single DC nodes $v^1 \in \mathcal{V}^{\rm dc}$ are considered, which we feel is sufficient for illustrating the opportunities for enhancing resilience and reducing costs in distributed MECs. However, more elaborate models can easily be incorporated into the optimization formulation in the next section.
It is assumed in this model that, if an entire DC becomes unavailable, additional spare capacity $c^{\rm dc-sp-tot}_{v^2}$ must be reserved at one or more secondary DCs $v^2$ so that users can quickly reestablish their sessions. Note that the superscripts 1 and 2 represent primary and secondary serving DC allocations. For latency-constrained MEC flows, the secondary DC(s) must also be within the serving area of the user's serving BS. Furthermore, there must be sufficient spare capacity on links along paths to the secondary DC(s) to accommodate the failover traffic. 

We point out that this model is, fundamentally, a departure from the `1+1' redundancy practiced by operators currently; instead of statically assigning flows to statically-provisioned primary and secondary NE nodes, mobile cloud controllers can dynamically reserve resources at various DC sites based on the current load and re-route/re-distribute flows amongst this set of candidate failover sites as needed.

\section{Optimization Formulation and Algorithm}
\label{sec:mec_opt}

We now formally state the minimum-cost MEC resource allocation problem as a Mixed Integer Linear Program (MILP), which incorporates the aforementioned costs and constraints. The objective function can be expressed as
\begin{align}
\label{eq:obj_func}
\textit{MEC-MILP}:   & \nonumber \\
\underset{x_{i,v^1}}{\min} & \sum_{v\in\mathcal{V}^{\rm dc}} \Big( \sum_{i \in SA(v)} h_i x_{i,v} b^{\rm dc} + c^{\rm dc-sp-tot}_{v} + \nonumber \\ 
& \sum_{p \in P_{(i,v)}} \sum_{e \in p} h_{i,p} b^{\rm link}_e \Big) + \nonumber \\
& \sum_{e \in \mathcal{E}^{\rm dc}} c^{link-sp-tot}_e b^{\rm link}_e + \nonumber \\
& \sum_{i\in \mathcal{V}^{\rm bs}} \sum_{j \in \mathcal{N}(i)} \Lambda^{\rm sar}_{i,j} b^{\rm sar} 
\end{align}

In (\ref{eq:obj_func}), binary variables $x_{i,v^1}$ indicate the allocation of BS $i$ to primary DC $v^1$. To simplify the model, we allow the aggregate flow from each BS to be assigned to only one primary serving DC in its SA.
The first term of (\ref{eq:obj_func}) represents the cost of allocating capacity for the aggregate demand $h_i$ from each BS $i$ to each primary serving DC $v$, whereas the second term is the cost of spare capacity $c^{\rm dc-sp-tot}_{v}$ reserved at each $v$ needed to recover from a single primary DC failure. The third term is the total cost of link capacity needed to transport the demand along feasible path-links from $i$ to $v$ and the fourth term is the cost of spare capacity $c^{link-sp-tot}_e$ allocated on each link $e$. The final term is the total cost of SAR $\Lambda^{\rm sar}_{i,j}$ between BS $i$ and one of its neighbors $j$. We note that the time dependence of $h_i$, along with the HO/SAR variables, is dropped for brevity. 


The minimization is subject to a number of constraints, stated as follows.
\begin{enumerate}[i.]
\item \textit{Primary DC capacity constraint:} For traffic demand $h_i$ of BS $i$ to be served by a primary DC $v^1$, sufficient processing capacity must be provisioned at $v^1$. \label{eq:prim_demand}
\item \textit{Primary path capacity constraint:} There must be sufficient bandwidth along one or more feasible paths $p \in P_{(i,v^1)}$ to route traffic between $i$ and $v^1$. The aggregate demand $h_i$ split over each path, denoted $h_{i,p}$, must be less than the bandwidth of any path-link along $p$.
\item \textit{Redundant (spare) DC capacity constraint:} If $h_i$ Mbps of capacity is allocated at primary DC $v^1$, we must ensure an equal amount of spare capacity is reserved at other DCs $v^2$ in the service area of BS $i$ to recover from a failure of $v^1$.
Since multiple BS flows may be allocated to primary $v^1$ that could fail over to $v^2$, the total spare capacity at $v^2$ must be greater than the sum spare capacity contributed from all of these failover flows. \label{eq:dc_spare}
\item \textit{Total DC capacity constraint:} The sum of the primary and spare capacity allocated at each DC cannot exceed its maximum capacity.
\item \textit{Redundant (spare) path bandwidth constraint:} Spare bandwidth, equal to the spare capacity allocated for BS $i$ at secondary DC $v^2$, must also be reserved along paths $p \in P_{(i,v^2)}$ in order to transport failover traffic. \label{eq:path_spare}
\item \textit{Total path-link capacity constraint:} Like the total spare DC capacity, the total spare link capacity $c^{\rm link-sp-tot}_e$ allocated on $e$ must be at least as much as the sum of the path capacities for all BS flows, which failover from some $v^1$ to some $v^2$.
Also, the sum of the primary and spare capacity allocated on link $e$ must not exceed its maximum capacity. \label{eq:link_cap_tot}


\item \textit{Total SAR constraint:} 
In this model, SAR occurs whenever a user moves from a source BS $i$ to a target BS $j$, where $x_{i,v} \ne x_{j,v}$ (i.e., the serving DC of the target BS differs from that of the source BS). Hence, the SAR between $i$ and $j$ can be written as:
\begin{align}
\Lambda^{\rm sar}_{i,j} = \lambda_{i,j} \Big( (x_{i,v_1} \oplus x_{j,v_1})~\lor~&(x_{i,v=2} \oplus x_{j,v_2})~\lor \nonumber\\ \label{eq:sar_con}
...~\lor~ (x_{i,v={N^{\rm dc}_i}} \oplus x_{j,v_{N^{\rm dc}_i}}) \Big)
\end{align}
where $N^{\rm dc}_i = \lvert DC(i) \rvert$ is the number of candidate serving DCs for BS $i$ and $\oplus$ is the exclusive-or (XOR) operation. Since, for each $i$, there will be only one non-zero $x_{i,v}$ variable in (\ref{eq:sar_con}), XORing the binary assignment variables for the neighboring BSs ensures that the SAR $\Lambda^{\rm sar}_{i,j}$ is equal to the total HO rate $\lambda_{i,j}$ only when $i$ and $j$ are assigned to different DCs and is equal to 0 otherwise. We note that it is possible to express (\ref{eq:sar_con}) in a form compatible with MILP standard form, although for brevity the exact constraints are not expressed here.
\end{enumerate}

\subsection{Heuristics for Large-Scale MEC Optimization} 
\label{sec:mec_algorithm}
Considering the large number of variables and constraints that would be required -- with one binary variable $x_{i,v}$ for potentially every BS-DC pair -- it is not feasible to solve the \textit{MEC-MILP} problem using global optimization techniques such as branch-and-bound even for medium-sized networks. Therefore, we now propose a number of relaxations and heuristic techniques that can be used to simplify the problem and allow for efficient optimization of large-scale MEC networks. 

\subsubsection{Initial Isolating K-cut}
\label{sec:kcut}
First, the \textit{handover graph} $\mathcal{G^{\rm ho}} = (\mathcal{V^{\rm bs}},\mathcal{E^{\rm ho}})$ is generated by taking the access/backhaul graph $\mathcal{G}^{\rm bs}$ and adding edges between each adjacent BS, which are weighted by the HO rates $\lambda_{i,j}$. The backhaul edge weights between each BS $i$ and its adjacent leaf DCs are assigned the same arbitrarily-large weight.\footnote{The weight on backhaul edges must be large enough to guarantee that the $K$-cut does not produce regions containing only isolated DCs with zero BSs.} Then, any of the well-known approximation schemes (see \cite{KernighanLin:70,KarypisKcut:98}) may be employed for finding the minimum \textit{$K$-cut}, which isolates each of the $K$ DCs $v \in \mathcal{V}^{\rm dc}$ into separate regions or \textit{clusters}, each cluster containing a single DC and one or more BSs. For this initial clustering, all leaf DCs are utilized as serving DCs (i.e., the network is fully-distributed) and the cost of HOs/SAR between clusters is within a factor of $2 - \frac{2}{K}$ from the minimum. 


\subsubsection{Initial BS-to-Leaf DC Assignment and LP Relaxation}
\label{sec:lp_assign}
Following the initial $K$-cut, the \textit{MEC-MILP} problem can then be simplified by aggregating the demand from all BSs within each DC cluster. First, the binary $x_{i,v^1}$ variables in the first term of (\ref{eq:obj_func}) and in constraints (\ref{eq:prim_demand}) and (\ref{eq:dc_spare}) are replaced with $x_{l,v}$, where $l$ denotes the leaf DC initially serving the cluster (that is, the first hop from the BSs of the cluster). Similarly, $i \in \mathcal{V}^{\rm bs}$ is replaced with $l \in \mathcal{V}^{\rm dc}$ in these and subsequent constraints (\ref{eq:prim_demand}) through (\ref{eq:link_cap_tot}). Constraints are then added to assign the aggregate demand $h_l = \sum_{i \in BS(l)} h_i$ (where $BS(l)$ is the set of BSs rooted at leaf DC $l$) to the local leaf DC $l$ representing each of the $K$ clusters by forcing $x_{l,l} = 1$ and $x_{l,v} = 0, \forall v \ne l$. This step reduces the number of constraints considerably and results in the problem being converted into a linear program \textit{MEC-LP}, which can easily be handled by commercial solvers. We abuse notation somewhat by interpreting $v \in DC(l)$ or $l \in SA(v)$ to henceforth mean that the leaf DC $l$ is within the path latency budget of DC $v$. 



\subsubsection{Greedy Pairwise Cluster Merge}
The fully-distributed case resulting from the initial clustering and assignment of BSs to serving DCs may not be the optimal, min-cost assignment, since the total amount of SAR between the $K$ DC clusters may still be high. A procedure called \textit{greedy pairwise cluster merge} thus needs to be performed to iteratively reduce the SAR cost. For each pair of neighboring clusters, represented by their serving DCs $v$ and $u$, we compute the \textit{pseudocost} $\Gamma_{v,u}$, which is defined as the ratio of the total SAR between DCs $v$ and $u$ to the total primary DC and path-link cost associated with routing demand to $v$. 


In each iteration, the cluster pair $(v,u)$ with the largest $\Gamma_{v,u}$ is selected to be merged together. The merge operation involves allocating all of the demand from DC $v$ to $u$, or visa-versa, depending on which assignment results in the lowest value of (\ref{eq:obj_func}) (after re-computing the solution to \textit{MEC-LP} for the updated cluster assignment). The total inter-cluster SAR rate 

\begin{equation}
\Lambda^{\rm sar}_{v,u} = \sum_{i \in SA(v)} \sum_{j \in SA(u)} \Lambda^{\rm sar}_{i,j}, ~~~ i,j \in \mathcal{V}^{\rm bs}
\end{equation}
between $v$ and $u$ thus becomes 0.
The underlying intuition behind this operation is that merging this pair will reduce SAR between the two former clusters while not shifting a large amount of demand to the new serving DC. Although the total cost of primary DC capacity is invariant, since all primary demand must be served by some DC, it is still best to avoid concentrating too much capacity at any one DC, as this would likely result in more overall \textit{redundant} capacity being required at other DCs.

\section{Simulation Setup and Results}



To illustrate the key benefits and limitations of distributed, MEC-based core networks enabled by SDN/NFV, we apply the optimization from the previous section to a real-world cellular network. 
Our general approach is to use the topology,  data traffic and handover statistics of a major cellular operator's network as input to several simulations where we consider different combinations of costs and parameters. Specifically, we have obtained a subset of the topology containing 2 major datacenters, 35 central offices/switching centers, 100 backbone links and 2000 4G eNodeB base stations from a dense 160 km\textsuperscript{2} metropolitan area, in addition to the hourly data traffic volume per eNodeB and inter-eNodeB handover rates over a single day.  

In the operator's current network, only 2 of the selected sites are DCs where EPC entities are located. However, it is assumed for the sake of this analysis that the other 35 COs have been converted into MEC-enabled DCs, resulting in a total of 37 DC sites capable of hosting core VNFs and cloud services. 

While it is assumed the operator will be able to assign some monetary value to each of the cost parameters, the actual costs are not available to us. Therefore, our approach is to perform a sensitivity analysis in which we vary the input costs and find the minimum-cost allocation of resources and routing of flows for each set of costs. 

\subsection{Impact of SAR Cost on Serving DC Distribution}
\label{sec:sar_test}
A key relationship we wish to shed light on is how the cost of SAR influences the degree to which the operator will want to distribute MEC functions and services when their goal is to minimize the total network cost. Toward this end, we vary the SAR cost $b^{\rm sar}$ while keeping the DC and link capacity costs fixed (in this case, the $b^{\rm dc}$ and $b^{\rm link}$ constants are arbitrarily set to \$1 per Mb). The cost $b^{\rm sar}$ is increased from 0 until the point is reached where the min-cost solution involves all clusters being merged into one cluster served by a single DC (in which case, the total SAR rate is 0). Cases where a maximum round-trip path delay of 10 ms and 1 ms is enforced between BSs and serving DCs are considered separately. DC and link capacities are assumed to be infinite (i.e., they are uncapacitated). 



With the SAR cost set to 0, all BS demand is assigned to the local leaf DC in each of the original 37 clusters resulting from the initial isolating cut. As shown in Figure~\ref{fig:num_dc}, we see that all DCs are being utilized for serving primary as well as failover traffic. This also results in the maximum SAR volume (for the initial clustering computed from the minimum 37-cut of $\mathcal{G^{\rm ho}}$).
Here the optimal assignment is dominated by the fixed DC and link capacity costs. Intuitively, as the SAR cost increases, the min-cost solution will tend to include fewer and fewer serving DCs, as the cost of migration begins to dominate the effect that distribution has on reducing the overall DC cost, as well as the additional link cost of transporting traffic away from leaf DCs to more centralized DCs. 
In Figure~\ref{fig:sar_cost_10ms}, we observe how, as $b^{\rm sar}$ (or, equivalently, the ratio $\frac{b^{\rm sar}}{b^{\rm dc}}$) increases beyond $10^3$, the link cost begins to contribute more to the total cost since the optimal solution favors a smaller set of more centralized serving DCs and traffic to/from the access network must be transported over greater distances (again noting that link cost is proportional to distance). When $b^{\rm sar}$ increases beyond $10^5$, we have the other extreme case where only 1 primary and 1 secondary serving DC is selected in the 10 ms RTT case. At this point, the total SAR no longer increases, as there are no longer any inter-DC migrations. For the 1 ms RTT case, however, it is not possible for any one DC to serve the entire region. In fact, it is found that 4 is the minimum number of DCs required to cover the entire set of BSs for this region within the 1 ms latency budget.

\subsection{Relationship of Distribution to Resilience}
\begin{figure}
	\centering
	\begin{subfigure}[b]{0.48\textwidth}
		\centering
		\includegraphics[trim={2.5cm 7cm 2.5cm 7cm},width=.7\textwidth]{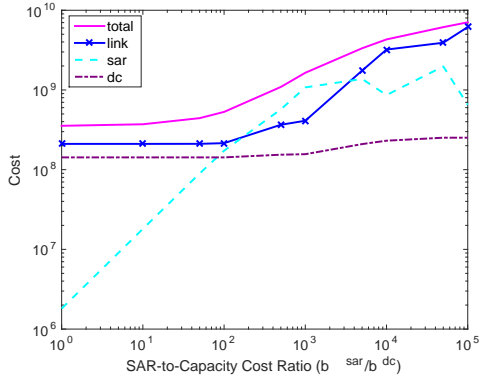}
		\caption{Link, DC and SAR costs as functions of SAR cost for 10 ms case}
		\label{fig:sar_cost_10ms}
	\end{subfigure}
	~
	\begin{subfigure}[b]{0.48\textwidth}
		\centering
		\includegraphics[trim={2.5cm 7cm 2.5cm 7cm},width=.7\textwidth]{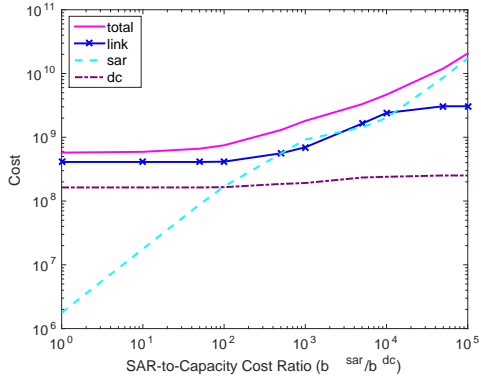}
		\caption{Link, DC and SAR costs as functions of SAR cost for 1 ms case}
		\label{fig:sar_cost_1ms}
	\end{subfigure}	
	~
	\begin{subfigure}[b]{0.48\textwidth}
		\centering
		\includegraphics[trim={2.5cm 7cm 2.5cm 7cm},width=.7\textwidth]{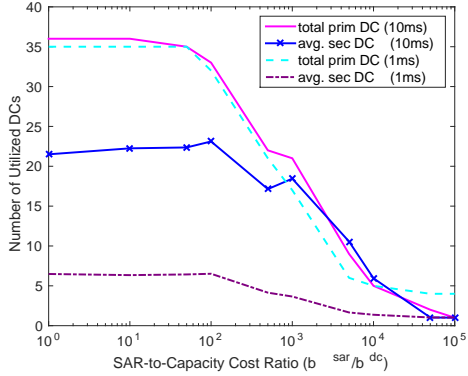}
		\caption{Number of utilized primary and secondary DCs as a function of SAR cost}
		\label{fig:num_dc}
	\end{subfigure}
	~
	\begin{subfigure}[b]{0.48\textwidth}
		\centering
		\includegraphics[trim={2.5cm 7cm 2.5cm 7cm},width=.7\textwidth]{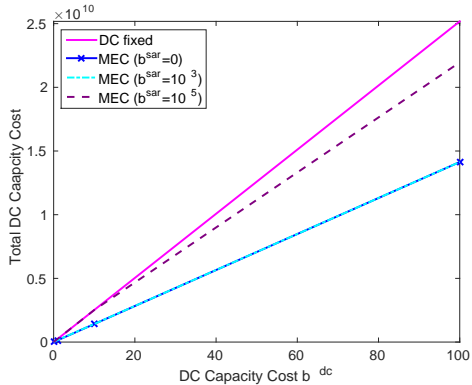}
		\caption{Total DC cost for fixed DC placement vs. optimized MEC placement}
		\label{fig:dc_cost_ndc_vs_mec}
	\end{subfigure}
	\caption{Experimental Results for 37-node Topology}
\end{figure}

In Figure~\ref{fig:dc_cost_ndc_vs_mec} we see how the total DC cost increases as a function of the cost of DC processing capacity. The ``DC fixed'' curve corresponds to current topology with only 2 fixed DC sites serving the whole region. In this case, we see that the DC cost scales linearly with a factor of 1 relative to the capacity cost. Each of the ``MEC'' curves correspond to the result of the MEC optimization when run on all 37 sites for 3 selected SAR costs and with a latency constraint of 10 ms. One notices that, for each of the MEC results, the DC cost scales with a factor of less than 1, since the load is spread out and less redundant capacity is required across the set of DCs.  In the case where the SAR $b^{\rm sar}= 0$, 75\% more DC capacity is needed under the current, centralized CN topology and 1+1 redundancy model.

Also, as previously noted, with only the 2 current DC locations, not all demand can be serviced under the 1 ms latency constraint. In fact, only 60\% of the total traffic for the region is within 1 ms of these sites. In a separate study, it was determined that, by placing around 200 additional DCs located near major metropolitan centers in the US, 90\% of all BS sites would be within 1 ms round-trip latency of the closest DC.
If we neglect the SAR cost entirely and perform the optimization on a network with these 200 DCs (in addition to the current national DC sites, of which there are on the order of 10), we find that a 20\% reduction in redundancy is possible compared to the current deployment.


In summary, the fundamental tradeoff to be understood is that higher migration cost encourages more centralized topologies (which may be constrained by latency requirements), while costs of (i) redundant capacity and (ii) bandwidth on backbone links drives the architecture towards a more distributed setting. 

\subsection{Performance of Greedy Heuristic}
For a network of this size, it is possible to solve a modified \textit{MEC-MILP} problem, which results from performing the initial K-cut and assignment procedures. That is, after the 37-cut is found, the traffic from each cluster is aggregated at the local cluster leaf DC. The globally-optimal solution can then be found using a commercial MILP solver in a reasonable amount of time. 



It is seen that, for the 10 ms case, the total cost of the greedy method is exactly equal to the optimal MILP case for every value of $b^{\rm sar}$ except for the maximum of $10^5$, at which point it is only 5\% off from optimal. However, for the 1 ms case, the greedy and optimal solutions begin to diverge at values of $b^{\rm sar} = 10^4$ and above and is off from the optimal by 50\% in the worst-case. This result is reasonable, as in the 10 ms case, there are more degrees of freedom for merging clusters, since it is less likely that merging a cluster pair would cause the latency constraint of users within one of the original clusters to be violated. In the 1 ms case, on the other hand, there are fewer possibilities for performing valid mergers, which forces the greedy heuristic to settle on a locally-optimal point.

The greedy heuristic also terminates in less than 5\% of the time, on average, taken by the MILP solver (provided by the MATLAB\textsuperscript{TM} Optimization Toolbox) for the 37-DC network. We anecdotally find that, for larger networks, the MILP optimizer often does not terminate after an hour or more, making it impractical for periodic resource allocation in large-scale MECs.


\section{Conclusions and Future Work}

The availability of virtualized network functions in conjunction with intelligent routing via SDN 
gives carriers a path to a distributed and more flexibile architecture that can ``breathe'' based on demand and enhance resiliency, resulting in significant cost efficiencies. However, cellular networks involve challenges such as user mobility and diverse signaling/control plane requirements, which need to be considered when designing and provisioning such a virtualized platform. In this work, we have taken a generic architectural
model that accounts for constraints introduced by mobility and proposed approaches for provisioning the data plane for such MEC-based cellular networks that can improve resiliency, reduce latency {\em and} limit mobility-based service disruptions.  Our planned next steps are two-fold: refine our provisioning approach to account for time-varying traffic patterns/surges as well as architect and implement a design that can leverage the methods introduced in this paper for online resource allocation and routing. The former presents interesting challenges such as deciding how much to reconfigure the network in response to traffic fluctuations (so as to prevent excessive churn of states), while the latter is of course critical to implementing these designs in practice and evaluating their efficacy.

\bibliographystyle{IEEEtran}
\bibliography{bibl}

\end{document}